\documentstyle[12pt,sprocl,epsfig]{article}
\begin{document}
\newcommand {\rch} {$r_{ch.}$}
\newcommand {\epem}     {e$^+$e$^-$}
\newcommand {\durham}   {$k_\perp$}
\newcommand {\qq} {q$\overline{\mathrm{q}}$}
\newcommand {\gluglu} {gg}
\newcommand {\gincl} {g$_{\,\mathrm{incl.}}$}
\newcommand {\nchgg} 
   {$\langle n_{ch.} \rangle_{\mathrm{gg}\,\mathrm{events}}$}
\newcommand {\mngincl} {$\langle n_{ch.} 
      \rangle_{\mathrm{g}_{\,\mathrm{incl.}}}$}
\newcommand {\egincl} 
   {$\langle E \rangle_{\mathrm{g}_{\,\mathrm{incl.}}}$}
\newcommand {\mnuds} {$\langle n_{ch.} \rangle_{\mathrm{uds\,hemis.}}$}
\newcommand {\mnudsmod}
   {$\langle n_{ch.} \rangle_{\mathrm{uds\,hemis.}}^{39.2\,\mathrm{GeV}}$}
\newcommand {\mngl} {$\langle n \rangle_{\mathrm{gluon}}$}
\newcommand {\mnqu} {$\langle n \rangle_{\mathrm{quark}}$}
\newcommand {\ca} {C$_{\mathrm{A}}$}
\newcommand {\cf} {C$_{\mathrm{F}}$}
\newcommand {\ejet} {$E_{\,\mathrm{jet}}$}
\newcommand {\ecm} {\mbox{$E_{c.m.}$}}
\newcommand {\lmsnffive} {$\Lambda_{\overline{\mathrm{MS}}}^{(n_f=5)}$}
\begin{center}{\large  UNIVERSITY OF CALIFORNIA, RIVERSIDE
}\end{center}

\vspace*{.3cm}

\begin{tabbing}
\` UCRHEP-E181    \\
\` 1 December 1996   \\
\end{tabbing}

\centerline{\normalsize\bf TEST OF QCD ANALYTIC PREDICTIONS FOR}
\baselineskip=16pt
\centerline{\normalsize\bf GLUON AND QUARK JET 
DIFFERENCES~\footnote{Talk given at the XXVI International 
Symposium on Multiparticle Dynamics, Faro, PORTUGAL, 
September 1-5, 1996.}
}

\centerline{\footnotesize J. WILLIAM GARY}
\baselineskip=13pt
\centerline{\footnotesize\it Department of Physics, University of California,
}
\baselineskip=12pt
\centerline{\footnotesize\it Riverside, California 92521, USA }
\centerline{\footnotesize e-mail: bill.gary@cern.ch, 
william.gary@ucr.edu }

\section{Introduction and analysis method}

Differences between the properties of gluon and 
quark jets have been convincingly
established by experiments operating at
the {\epem} collider LEP at CERN~\cite{bib-jwgsl95}.
The most conclusive results
have been from so-called Y events~\cite{bib-opalstring91},
which are three jet events
in which the angle between
the highest energy jet and each of the two lower
energy jets is about 150$^\circ$.
Table~\ref{tab-rchyevents} summarizes the results
for the mean charged particle multiplicity ratio
between gluon and quark jets, {\rch},
found using Y events at LEP.
A 60\% variation in the value of {\rch} is observed
if the Cone or JADE-E0 jet finders are used
to identify the three jet events instead of the {\durham} 
(``Durham'') jet finder.

QCD analytic predictions exist for the ratio of
the mean particle multiplicity between gluon and quark 
jets~\cite{bib-qcdmult1}$^-$\cite{bib-qcdmult3}.
A quantitative test of these predictions
has not been possible, however,
because of differences between the theoretical and 
experimental definitions of the jets and event samples.
The analytic calculations employ definitions of the event
samples and jets which are entirely inclusive.
For the calculations,
two samples of events are chosen:
a sample of gluon-gluon {\gluglu} events
produced from a color singlet point source is used to define
the gluon jet properties and a sample of
quark-antiquark {\qq} events produced under the
same circumstances is used to define the quark jet properties.
The gluon and quark jet characteristics are given
by inclusive sums over the particles in these two samples.
Thus,
the theoretical results are not restricted to three-jet events
and do not employ a jet finder to assign particles to the jets,
in contrast to the Y event studies and other previous 
experimental studies of gluon and quark jet differences 
at high energy colliders.

The experimental difficulty in obtaining a jet definition
corresponding to the theoretical one lies in the
gluon jet sample,
since {\gluglu} production from a point source does not
occur naturally in {\epem} annihilations.
In contrast, the {\qq} sample employed by the theory is
the inclusive {\epem} multihadronic one and so 
is readily available.
In ref.~\cite{bib-jwg},
a method was proposed for LEP experiments
to identify gluon jets using an inclusive
definition similar to that used for the analytic calculations.
The method is based on rare events
of the type {\epem}$\rightarrow\,${\qq}$\,${\gincl}
in which the q and $\overline{\mathrm{q}}$ are
identified quark jets which appear
in the same hemisphere of an event.
The quantity {\gincl}, taken to be the gluon jet,
is defined by the sum of all particles observed in the
hemisphere opposite to that containing 
the q and~$\overline{\mathrm{q}}$.
In the limit that the q and~$\overline{\mathrm{q}}$ are collinear,
the gluon jet {\gincl} is produced under the same
conditions as the gluon jets in {\gluglu} events
from a color singlet point source.
The jets {\gincl} therefore correspond closely to 
single gluon jets in {\gluglu} events,
defined by dividing the {\gluglu} events in half
using the plane perpendicular to the principal event axis.

\begin{table}[t]
\centering
\begin{tabular}{|c|ccc|}
  \hline
   Experiment & {\durham} & Cone & JADE-E0 \\
  \hline
  \hline
    ALEPH~\cite{bib-aleph95} 
       & $1.19\pm0.04$ & ---           & --- \\
    DELPHI~\cite{bib-delphi95} 
       & $1.24\pm0.03$ & ---           & $1.37\pm0.04$ \\
    OPAL~\cite{bib-opal95}   
       & $1.25\pm0.04$ & $1.10\pm0.03$ & ---\\
  \hline
\end{tabular}
\caption{
LEP results for the charged particle multiplicity ratio
between gluon and quark jets, {\rch},
for Y events defined using the {\durham}, Cone
and JADE-E0 jet finders.
}
\label{tab-rchyevents}
\end{table}

\begin{figure}[t]
\begin{center}
\begin{tabular}{ll}
  \epsfig{file=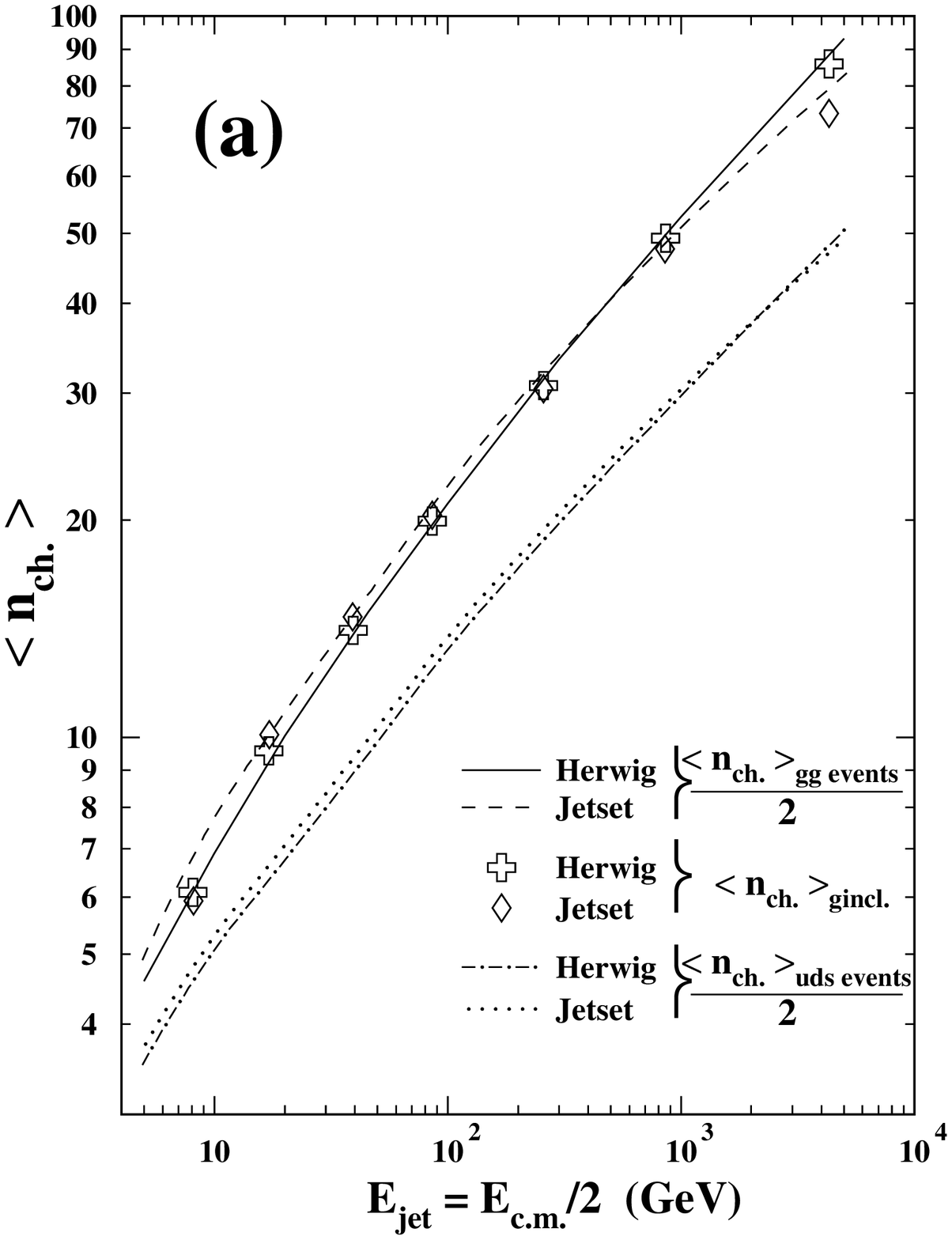,height=8cm} &
  \epsfig{file=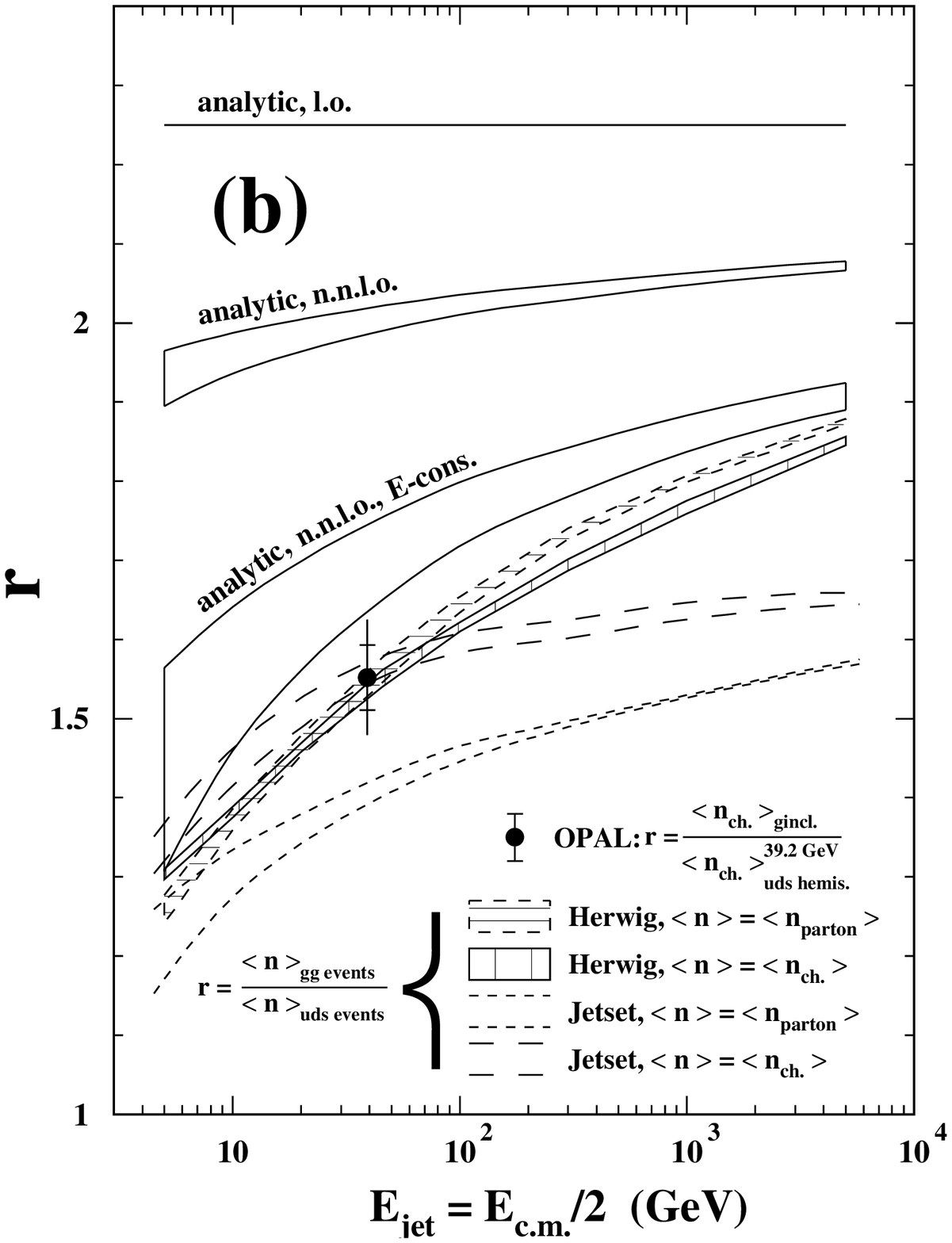,height=8cm} \\
\end{tabular}
\end{center}
\vspace*{0.8cm}
\caption{
(a)~Predictions of QCD Monte Carlos
for the mean charged particle multiplicity
in {\gluglu} event hemispheres and in {\epem} {\gincl} jets,
as a function of jet energy.
The corresponding predictions for uds {\qq} event
hemispheres are also shown.
(b)~Measured result for the multiplicity ratio {\rch}
between gluon and uds jets at {\ejet}=39~GeV
in comparison to analytic and Monte Carlo predictions
shown as a function of~{\ejet}.
}
\label{fig-methodtest}
\end{figure}

Although {\gluglu} events from a color singlet
point source are not produced in {\epem} annihilations,
they may be generated using QCD Monte Carlo programs
such as the Herwig~\cite{bib-herwig}
and Jetset~\cite{bib-jetset} parton shower models.
To illustrate that the definition of gluon jets
{\gincl} from the {\epem} events corresponds closely to that
employed by theory from the {\gluglu} events,
the Monte Carlo is used to
calculate the mean charged particle multiplicity
in {\gluglu} events, {\nchgg},
and the results are divided by two so that 
they correspond to a single hemisphere:
these results are shown as a function of the
jet energy E$_{\mathrm{jet}}$$\equiv$E$_{c.m.}/2$
by the solid and dashed lines 
in Fig.~\ref{fig-methodtest}(a).
Shown in comparison,
by the cross and diamond symbols,
are the results for {\gincl} jets from {\epem} events
generated using Herwig and Jetset, respectively.
The results for the {\gluglu} and {\gincl} samples 
are seen to agree well over the entire jet energy range
from 5~GeV to $5\,000$~GeV,
i.e. {\mngincl}$\,\approx\,\frac{1}{2}\,${\nchgg}
for both Herwig and Jetset.
This establishes that the properties of the {\gincl}
sample derived from the {\epem} data do indeed correspond
closely to those of {\gluglu} events generated from
a color singlet point source,
confirming the results of ref.~\cite{bib-jwg}.

Below,
the results of a recent study~\cite{bib-opalgincl}
which uses this analysis method are presented.
The data sample employed is that
of the OPAL detector at LEP.
Details of the event selection are
presented in ref.~\cite{bib-opalgincl}.
In total,
278 events are selected for the final gluon jet
{\gincl} event sample.
With the final cuts,
Jetset with detector simulation and the same selection
criteria as are applied to the data
predicts that both jets in the tagged hemisphere
are quark jets with $(83.0\pm 1.7)$\% probability,
where the uncertainty is statistical:
this is the estimated purity of the {\gincl} gluon jet sample.
The mean energy of the gluon jets, {\egincl},
is determined to be
{\egincl}$\,=39.2\pm 0.3\,\mathrm{(stat.)}
\pm 1.8\,\mathrm{(syst.)}$~GeV.
The uds quark jets in the study are defined inclusively
using particles observed in event hemispheres
opposite to those containing identified uds jets.
The selection criteria are described in ref.~\cite{bib-opalgincl}.
In total, $28\,007$ uds hemispheres are tagged.
The estimated uds purity of the sample,
obtained by treating Jetset events with detector simulation
in the same manner as the data, is $(93.2\pm 0.2)$\%,
where the uncertainty is statistical.
The energy of the uds jets is given by the beam
energy, 45.6~GeV.

\section{Results}

The mean charged particle multiplicity value
measured for the gluon jets is
{\mngincl}$=14.63\pm 0.38\,\mathrm{(stat.)}
\pm 0.59\,\mathrm{(syst.)}$.
The corresponding result for the uds jets is
{\mnuds}$=10.05\pm 0.04\,\mathrm{(stat.)}
\pm 0.23\,\mathrm{(syst.)}$.
Before forming the ratio between the gluon
and uds jet measurements,
it is necessary to account for the different energies
of the two samples:
the gluon jets have a mean energy of 39.2~GeV
while the uds jets have a mean energy of 45.6~GeV.
To account for the larger energy of the uds jets,
the QCD analytic formula for the evolution of
the mean event multiplicity in
{\epem} annihilations~\cite{bib-qcdnecm}
is employed.
Assuming $n_f$=$5$,
the QCD evolution formula predicts the mean multiplicity
in 78.4~GeV events to be $(6.2\pm 0.4)$\% smaller than
in 91.2~GeV events,
where the uncertainty results from the maximum variation 
found by using the jet energies (39.2 and 45.6~GeV)
rather than the event energies,
$n_f$=$3$ rather than $n_f$=$5$,
and varying the value of $\alpha_S$
within its allowed range~\cite{bib-pdg95}.
Applying a multiplicative correction of $0.938\pm 0.004$
to the 45.6~GeV uds jet measurement presented above yields
{\mnudsmod}=$9.43\pm 0.06\,\mathrm{(stat.)}\pm 0.22\,\mathrm{(syst.)}$
for the mean multiplicity of 39.2~GeV uds jets.

The result for the multiplicity ratio {\rch} between
39~GeV gluon and quark jets is therefore found to be:
\begin{equation}
\label{eq-rdef2}
   r_{ch.} \equiv
   \frac{ \langle n_{ch.} \rangle_{g_{\,\mathrm{incl.}}} }
   { \langle n_{ch.} \rangle_{\mathrm{uds\,\,hemis.}}^{39.2\,\mathrm{GeV}} }
   =
   1.552\pm 0.041\,\mathrm{(stat.)}\pm 0.061\,\mathrm{(syst.)}
    \;\;\;\; .
\end{equation}
The systematic uncertainty is discussed
in ref.~\cite{bib-opalgincl}.
It is to be noted that one of the systematic variations
consists in using the Cone or JADE-E0 jet finders
to tag quark jets for the {\gincl} identification,
rather than the {\durham} jet finder:
only a small change of $\Delta${\rch}$\approx$0.02
occurs as a result of this variation.
Thus, the result for {\rch} changes by only about 
4\% if different jet finders are employed for the analysis,
in contrast to the Y events for which a difference
of about 60\% is observed (Table~\ref{tab-rchyevents}).
This emphasizes that the result (\ref{eq-rdef2}) is only
weakly dependent on a jet finding algorithm.
This measurement of {\rch} is shown by the solid point in
Fig.~\ref{fig-methodtest}(b).
The experimental statistical uncertainty is indicated
by the small horizontal bars.

Various analytic results exist for the ratio
$r\equiv\,${\mngl}/{\mnqu} of the mean number of
partons in a gluon jet to that in a quark jet.
The original results~\cite{bib-qcdmult1},
valid to leading order, predict $r$
to be $r=\,${\ca}/{\cf}$=9/4$.
Later, higher order corrections valid to the
next-to-next-to-leading order were
found to reduce this result by about 10\%~\cite{bib-qcdmult2}.
These results do not incorporate energy
conservation into the quark and gluon branching processes.
Recently,
$r$ has been calculated including not only the
next-to-next-to-leading order terms but also
energy conservation~\cite{bib-qcdmult3},
and is found to be reduced yet further in magnitude.
Momentum conservation is not included in this latter calculation,
however:
therefore energy-momentum conservation is only approximate.
The analytic results for $r$,
valid to leading order (l.o.),
to next-to-next to leading order (n.n.l.o.),
and including approximate energy-momentum conservation
(n.n.l.o., E-cons.) are shown
in Fig.~\ref{fig-methodtest}(b) 
as a function of {\ejet}$\,=\,${\ecm}/2.
For the evaluation of the strong coupling constant, $\alpha_S$,
the values $n_f$=$5$ and {\lmsnffive}=$0.209$~GeV~\cite{bib-pdg95}
have been used,
where $n_f$ is the number of active quark flavors.
The results for the two n.n.l.o. calculations
are shown as bands:
the upper edges of the bands show the results if the 
energy scale used to evaluate $\alpha_S$ is taken to be {\ecm},
while the lower edges show the results
if this energy scale is taken to be~{\ecm}/4.
The widths of the bands therefore indicate the level of
uncertainty associated with the ambiguity of
the energy scale at which to evaluate~$\alpha_S$.
The predictions of Herwig and Jetset at both the
parton and hadron levels are included in
Fig.~\ref{fig-methodtest}(b).
The width of the bands for the Monte Carlo results
shows the variation which occurs
if the shower cutoff parameters are allowed
to vary within their allowed ranges~\cite{bib-opalgincl}.

For jet energies of 39~GeV,
the n.n.l.o. calculation which incorporates
energy conservation~\cite{bib-qcdmult3}
predicts values of $r$ between 1.83
(if $n_f$=$3$ and the energy scale of $\alpha_S$ is~{\ecm})
and 1.64
(if $n_f$=$5$ and the energy scale is~{\ecm}/4):
this last value is only slightly above
the measured result given above in relation~(\ref{eq-rdef2}).
The analytic result is valid for quarks and gluons
while the measurement refers to charged hadrons.
Jetset predicts a hadronization correction for~{\rch},
defined by the ratio of the parton to
hadron level curves in Fig.~\ref{fig-methodtest}(b)
(for {\ejet}$\,=39$~GeV), of~0.91.
The corresponding prediction from Herwig is~1.02.
In this sense,
the hadronization correction can be estimated to
be about unity
and to have an uncertainty of about~10\%.
Given the ambiguities of the energy scale
at which to evaluate $\alpha_S$,
of the number of active flavors $n_f$,
of the hadronization correction,
and due to the approximate nature by which energy-momentum
conservation is included,
is it seen that the analytic calculation of
Dremin, Hwa and Nechitailo~\cite{bib-qcdmult3}
is in general agreement with the measurement.
In contrast,
the analytic results which do not incorporate
energy \mbox{conservation~\cite{{bib-qcdmult1},{bib-qcdmult2}}}
are seen to be in clear disagreement with this measurement,
\mbox{even considering the theoretical ambiguities.}

\section{ Summary }

The first quantitative test of
analytic predictions for the ratio of the mean
particle multiplicity between gluon and quark jets
has been presented.
The technique that allows this test is the
identification of gluon jets in an inclusive manner,
by all the particles observed in the event hemisphere
opposite to a hemisphere containing a
tagged quark and antiquark jet~\cite{bib-jwg}.
The resulting definition is in close correspondence
to the definition of gluon jets in analytic calculations,
for the first time in the analysis of high energy {\epem} data.
In this study,
the gluon jet measurement,
valid for 39~GeV jets,
is compared to the corresponding measurement from
light quark (uds) jets,
which has also been defined inclusively.

The result for the ratio~{\rch},
the mean charged particle multiplicity
of gluon jets divided by the corresponding value
for uds quark jets, is
$r_{ch.}$=$1.552\pm 0.041\,\mathrm{(stat.)}$
$\pm 0.061\,\mathrm{(syst.)}$.
This result is substantially smaller than
the predictions of analytic calculations which do not
include energy conservation in the parton branchings.
A recent analytic calculation~\cite{bib-qcdmult3}
which incorporates approximate energy-momentum 
conservation predicts a parton level
multiplicity difference between 39~GeV gluon and quark jets
in the range from about 1.64 to~1.83,
depending on the choice for the energy scale $Q$
at which the strong coupling constant $\alpha_S(Q)$ is evaluated
and on the number of active quark flavors, $n_f$.
This latter prediction is in overall agreement with the measurement,
given the uncertainties
due to the approximate nature of energy-momentum 
conservation in the calculation,
missing higher order terms,
the energy scale, 
the number of active flavors $n_f$, and hadronization.

\end{document}